\documentclass[english,aps,prl,showpacs,preprintnumbers,amsmath,amssymb,twocolumn]{revtex4-1}
\usepackage[T1]{fontenc}
\usepackage[latin9]{inputenc}
\usepackage{amsmath}
\usepackage{amssymb}
\usepackage{graphicx}

\makeatletter

\@ifundefined{textcolor}{}
{%
 \definecolor{BLACK}{gray}{0}
 \definecolor{WHITE}{gray}{1}
 \definecolor{RED}{rgb}{1,0,0}
 \definecolor{GREEN}{rgb}{0,1,0}
 \definecolor{BLUE}{rgb}{0,0,1}
 \definecolor{CYAN}{cmyk}{1,0,0,0}
 \definecolor{MAGENTA}{cmyk}{0,1,0,0}
 \definecolor{YELLOW}{cmyk}{0,0,1,0}
 }

\def\gev{\; \hbox{GeV}}
\def\to{\rightarrow}

\usepackage{amsfonts}
\usepackage{epsfig}\usepackage{dcolumn}
\usepackage{bm}
\date{\today}

\makeatother

\usepackage{babel}
\begin{document}

\title{Is There a Hidden Principle in the Higgs Boson Decay to Photons?
\bigskip{}
}

\author{Alexandre Alves$^{1}$,  E. Ramirez Barreto$^{2}$, A. G. Dias$^{2}$ \medskip{}
}

\affiliation{$^{1}$Universidade Federal de S\~ao Paulo, UNIFESP,  Dep. de Ci\^encias Exatas e da Terra,  Diadema - SP 09972-270 Brasil, \\
$^{2}$Universidade Federal do ABC, UFABC, Centro de Ci\^encias Naturais e Humanas,  Santo Andr\'e - SP, 09210-170, Brasil.}

\begin{abstract}
It is remarkable that the measured Higgs boson mass is so close to the value which maximizes the Higgs decay rate to photons as predicted by the Standard Model. In this work we explore the consequences to assume that an $\sim 125$ GeV Higgs boson mass is not accidental, but fixed by some fundamental principle that enforces it to maximize its decay rate into photons. The principle is motivated by the evidence that only a very small volume of the parameters space of the Standard Model, which contains their measured values, could lead to a maximal Higgs boson with that mass. If the principle actually holds, several Standard Model features get fixed, as the number of fermion families, quark colors, and the CP nature of the new boson, for example. We also illustrate how such principle can place strong bounds on new physics scenarios as a Higgs dark portal model and a Two Higgs Doublet Model.
\end{abstract}
\maketitle

\section{Introduction}
\label{intro}

In spite of the fact that a light Higgs boson with a mass close to 114 GeV is required to fit the precision electroweak data, and apart from bounds imposed by perturbative unitarity, Higgs self-couplings triviality and scalar potential stability, which range from 100 GeV to 1 TeV approximately, the Higgs boson mass is a free parameter within the framework of the Standard Model (SM).

Additionally, the decay pattern of the SM Higgs boson is dictated by an intricate interplay between all the decay channels accessible to it. Changing the masses of the gauge bosons and fermions, or their couplings to the Higgs boson, alter significantly the decay branching fractions of the Higgs boson.

In special, the branching ratio to photons, which is very small ($\sim 10^{-3}$), depends upon several theory parameters and masses. Changing these parameters also alter the slope of the Higgs to photons branching ratio curve substantially. Given that sensitivity, it is remarkable that within the whole range of Higgs boson masses allowed by general theoretical principles, one that gets so close to the point of maximum decay rate to photons within the SM has been experimentally observed. If that was not the case, perhaps the LHC runs at 7 and 8 TeV would not have observed a Higgs boson in the $\gamma\gamma$ channel until this moment. For example, a Higgs boson with a mass larger than $\sim 145$ GeV has a $\sigma_{ggF}\times BR(h\to\gamma\gamma)$ which is 50\% smaller than a 125 GeV Higgs at least~\cite{hwg}.

In this work, we are going to show that the Higgs boson mass that leads to the maximal branching fraction to photon pairs is compatible to the combined central value obtained by the ATLAS~\cite{atlas14} and CMS~\cite{cms14} collaborations of $\sim 125.2$ GeV. We also show that varying some of the most relevant input parameters, only a small volume of the parameters space, in which the measured parameters lie, can lead to a maximal Higgs boson compatible with the experiments.  Motivated by the observation of such high sensitivity to the theory parameters and the remarkable coincidence between the measured and maximal Higgs mass, we propose that the value of the SM Higgs boson mass is fixed by some fundamental principle enforcing a maximal branching fraction to diphotons.

Moreover, a maximal  $\sim 125$ GeV Higgs fixes the inner structure of the SM, its number of fermion generations and quark colors, and coupling patterns, for example. It also excludes fermiophobia, gaugephobia, favors an unmixed real scalar Higgs scalar, and is capable to determine the masses and couplings to the Higgs of many SM particles to a high accuracy.

 If the Higgs boson mass, or any other scalar participating in the electroweak symmetry breaking (EWSB) actually, follows from the principle requirement we are assuming, not only the SM mass spectrum would get tightly constrained, but any new model contributing to the Higgs decays as well. Such a principle, then, would lead to far reaching consequences for model building. We illustrate the predictive power of the principle analyzing some consequences for a Higgs dark matter portal model and a Two Higgs Doublet Model (THDM).

By the way, we are going to show that the SM is finely tuned to give rise a maximal Higgs boson what might suggest an underlying unification principle like supersymmetry, which is much more natural from the point of view of higher corrections to the Higgs mass, for example.

We point out that a similar observation was made concerning the product of the most relevant Higgs branching fractions. Surprisingly an $\sim 125$ GeV Higgs boson corresponds to a maximum point of this product, which was hypothesized as being due a ``maximum opportunity'' underlying physical mechanism in Ref.~\cite{david}, an outcome resulting from the fact that the fermionic, bosonic and loop-mediated Higgs decays feature all strongly anti-correlated changes around $m_h \approx 1.5m_W$ due to their different $m_h$-power dependencies.

\section{Determining the Higgs boson mass}

First of all, we are going to show that requiring the Higgs mass to maximize the branching ratio to photons determines its mass to a high accuracy.

The Higgs branching ratio to photons involves its total width $\Gamma_{tot}=\Gamma_{\gamma\gamma}+\Gamma_R$ calculated as the sum of the loop induced partial decay width to photons $\Gamma_{\gamma\gamma}$~\cite{gunion,kniehl} plus the sum of all the partial widths from particles that the Higgs decays to, including 2, 3 and 4-body decays, $\Gamma_R$~\cite{djouadi}. The total width brings up sensitivity to the whole mass spectrum and couplings of the theory to the branching ratio.

The branching ratios of many SM particles exhibit maximum and minimum points as functions of the Higgs mass~\cite{djouadi}. We shall denote a Higgs mass corresponding to a point of maximum as $M_h$. Some of these maxima are correlated to the production threshold of the massive particles, as the top quark, $W$, and $Z$ bosons. In the case of decay to massless particles, the maximum is determined mainly by the behavior of all particles that compete in a certain mass range.

Concerning the Higgs branching ratios into photons, gluons and $\gamma Z$ cases, the maximum ratios occur in the mass region where the branching ratio to light fermions fall and the $WW^*$ and $ZZ^*$ rise. On the other hand, a heavy top quark decay does not influence the decay to photons too much, despite its importance in loop-induced decays of the Higgs boson.
\begin{center}
\begingroup
\begin{table}
\begin{tabular}{c|c|c|c|c|c|c|c}
\hline
\hline
channel & $b\overline{b}$ & $gg$ & $\gamma\gamma$ & $\gamma Z$ & $ZZ$ & $WW$ & $t\overline{t}$ \\
\hline
 $M_h$ & $29.4$ & $118.7$ & $125.4$ & $143.2$ & $149.3$ & $169.3$ & $484.1$\\
\hline
\hline
\end{tabular}
\caption{The Higgs boson mass in GeV that maximizes the branching ratio of various decay channels calculated from \texttt{HDECAY}~\cite{hdecay}. The parameters are taken from the Higgs Working Group report~\cite{hwg}, Eq.~\ref{pars}.}
\label{tab1}
\end{table}
\endgroup
\end{center}
%


We show in the table~\ref{tab1} the peak position of all channels that present a point of maximum in the SM. We computed all branching ratios with \texttt{HDECAY}~\cite{hdecay} which takes into account all the important NLO QCD and EW corrections to the partial widths plus 3 and 4 particle decay contributions from off-shell $W$ and $Z$ bosons. Higher order QCD and EW corrections to $\Gamma_{\gamma\gamma}$ are shown to be very small and are neglected~\cite{nnlo}.

The parameters with the highest impact on the maximal Higgs masses determination are those of Eq.~\ref{7dparams}, which are taken from the recommendations of the Higgs Working Group (HWG)~\cite{hwg}. The running of the quark masses are carried out in the NLO for the HWG parameters of Eq.~\ref{pars} below, as the quark masses from HWG correspond to the 1-loop pole masses of the charm, bottom, and top quarks.
\begin{eqnarray}
& & \!\!\!\!\!\!\!\!\!\!\!\!\!\!\!\Omega_{7D}\equiv\{m_W,m_Z,m_t,m_b,m_c,m_\tau,\alpha_S(m_Z)\}=\label{7dparams} \\
& & \!\!\!\!\!\!\!\!\!\!\!\!\!\!\!\{80.370,91.154,172.5,4.49,1.42,1.77,0.119\}_{HWG}\label{pars}
\end{eqnarray}
\begin{figure}
\begin{center}
\includegraphics[scale=0.5]{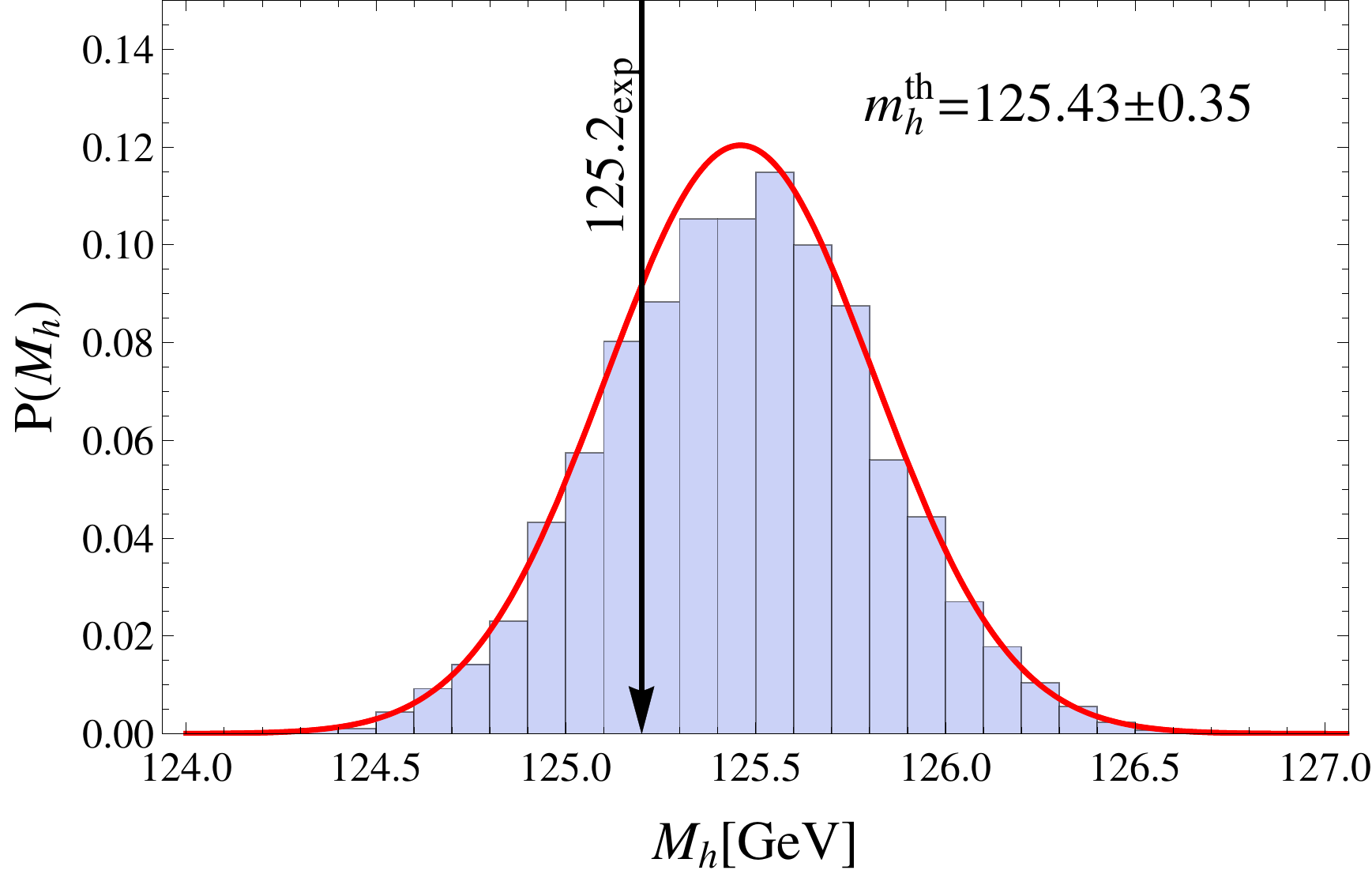}
\end{center}
\caption{ The SM Higgs boson mass determination requiring maximum decay rate to photons. The vertical arrow shows the combined CMS+ATLAS Higgs mass measurement.}
\label{fig:mh}
\end{figure}

 With these default values, the maximal Higgs boson mass, as can be read from Table~\ref{tab1}, is very close to the present measured central value of the Higgs boson from CMS~\cite{cms14} and ATLAS~\cite{atlas14} collaborations
\begin{eqnarray}
 m_h &=& (125.03\pm 0.30)\; \hbox{GeV} \;\;\;\; \hbox{CMS}\nonumber \\
 m_h &=& (125.36\pm 0.41)\; \hbox{GeV} \;\;\;\; \hbox{ATLAS}\nonumber \\
m_h & = & (125.17\pm 0.51)\; \hbox{GeV} \;\;\;\; \hbox{CMS+ATLAS}
\end{eqnarray}
where the combined experimental result is averaged by the inverse of the experimental errors.

There is an implicit uncertainty on these predictions from the uncertainties in the measured values of the input parameters though. We show in Fig.~\ref{fig:mh} the maximal Higgs mass distribution $P(M_h)=\int_{\Omega_{7D}}P(M_h|\vec{\theta})d^7\vec{\theta}$, obtained by taking random points $\vec{\theta}$ in the 7-Dimensional parameters space $\Omega_{7D}$ of Eq.~\ref{7dparams}, within their experimental errors as quoted by the HWG~\cite{hwg}. This is the set with the highest impact on the maximal Higgs masses determination. The Higgs mass determination requiring maximization of the branching fraction to photons is
\begin{equation}
m_h=(125.43\pm 0.35)\; \hbox{GeV}
\end{equation}

The difference is not larger than 0.5 experimental standard deviations (s.d.) and is perfectly compatible within one standard deviation to the combined experimental Higgs mass value.

We also observe that no other peak could possibly be associated to the measured Higgs mass as they are 20 s.d. away from the measured value at least. It is also interesting to notice that the nearest peaks to the experimental mass occur for decays involving at least one massless vector boson, that is it, $gg$, $\gamma\gamma$, and $Z \gamma$.

\section{Global sensitivity to the variations of parameters}

Given this remarkable coincidence, how likely is a Higgs boson mass near a point of maximum branching ratio considering all the SM particles? Let us consider a 2 GeV window around the Higgs mass, which corresponds approximately to the present 99\% confidence level (C.L.) mass region, and the  theoretically well motivated 100 GeV -- 1 TeV range for the Higgs mass.  In this mass region, the branching ratios to $\gamma\gamma$, $gg$, $Z\gamma$, $ZZ$, $WW$ and $t\bar{t}$ exhibit peaks. Assuming further a equiprobable distribution for the Higgs mass distribution, the chance that a Higgs boson mass lies so close to a branching ratio peak is around $\frac{[2\; \hbox{GeV}]\times 6}{[900\; \hbox{GeV}]}=1.3$\% given the present experimental resolution. This probability decreases as the experimental resolution of the measured mass improves. The probability of being near the $\gamma\gamma$ peak only drops to $\approx 0.2$\%.
\begin{figure}
\begin{center}
\includegraphics[scale=0.55]{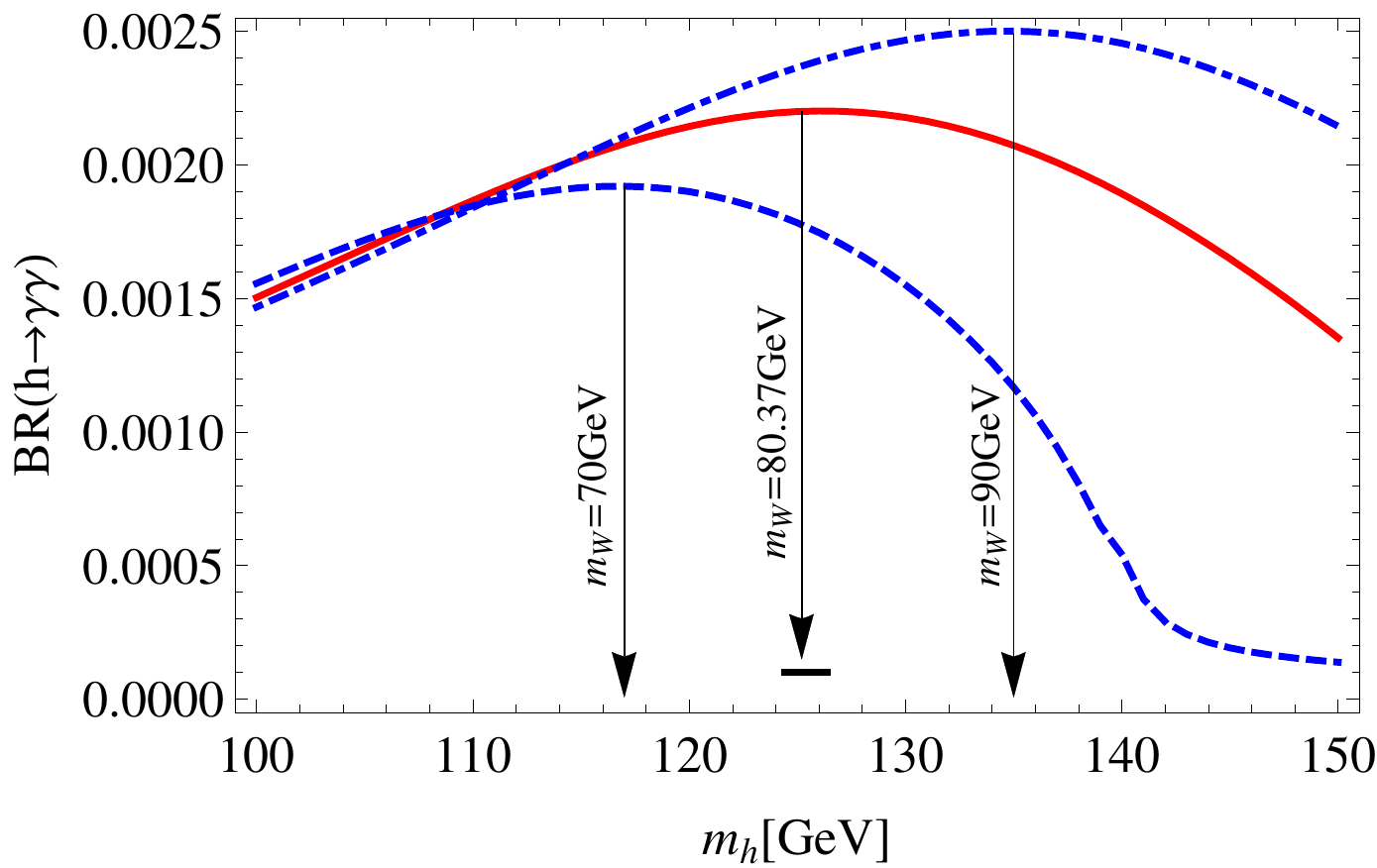}
\end{center}
\caption{Variation of the peak position of $BR(h\to\gamma\gamma)$ as a function of the $W$ boson mass. We also show the 95\%CL experimental Higgs mass region in the lower bar. The $Z$ and $W$ masses are related by $m_W=c_W m_Z$ in these plots. Their total widths were also updated accordingly.}
\label{fig:mwvar}
\end{figure}

In a broader sense, the probability that the branching ratio to photons reaches a maximum in the vicinity of a given Higgs mass should be calculated taking into account also the variation of all relevant parameters. This is the case concerning the maximal Higgs, once the peak position is strongly dependent on the parameters of the SM. In Fig.~\ref{fig:mwvar} we show the variation in the slope of $BR(h\to\gamma\gamma)$ for three $W$ boson masses: $70$, $80.385$, and $90$ GeV, fixing all parameters to their measured central values but the $Z$ mass, which was kept related to the $W$ mass through $m_W=cos\theta_W\, m_Z$ ($\theta_W$ is the electroweak mixing angle angle). The $Z$ and $W$ widths were also changed accordingly. The data point represents the combined fit to the Higgs mass with $2\sigma$ error bands. It is remarkable the high sensitivity of the slope, and the peak position, against variations of the $W$ mass. In fact, there is an enormous sensitivity of the peak position against all the SM parameters of $\Omega_{7D}$ given in Eq.~\ref{7dparams}.

Similar situation we encounter in the problem of extreme sensitivity of the Higgs mass up to the cutoff scale the SM is valid due radiative corrections. Even the minimal supersymmetric standard model (MSSM) now seems to be, in a much lesser extent tough, fine tuned to give rise a Higgs boson of $\sim 125$ GeV. In fact, fine tuning can be intuitively linked to a measure of the chance that set of parameters conspire in such a manner a given observable, which depends upon those parameters, have a specific value. For example, a heavy MSSM spectrum seems to be increasingly finely tuned to be compatible to the Higgs and the $Z$ boson masses. In other words, it is not natural to conceive very unlikely scenarios guided solely by pure chances in order to understand natural phenomena.

The connection between the probability of a given scenario and a measure of fine tuning was proposed in Ref.~\cite{ft}. A more reliable fine tuning measure should be sensible to all parameters at the same time, and it should embody a global sensitivity which is not the case for the widely used Barbieri-Giudice measure~\cite{bg}, for example. Instead, the fine tuning measure of Ref.~\cite{ft} compares volumes of solutions in a given parameters space for two different scenarios and then computes its likelihood. The fine tuning measure is the inverse of this likelihood which is intuitively very plausible.

Adopting the fine tuning measure of Ref.~\cite{ft}, which we denote by $\Delta$, we estimate the chance to randomly choose a point in the reduced 4-Dimensional space $\Omega_{4D}=\{m_W,m_b,m_t,\alpha_S(m_Z)\}$ which leads to a maximal Higgs, in the interval of 99\% C.L. around the measured mass. We chose $\Omega_{4D}$ as we were interested only in on-shell top quark decays allowed in \texttt{HDECAY}~\cite{hdecay} and to keep control on the perturbative regime of the quark masses. This is an estimate of the lower bound to that probability tough. In fact, we checked that all the other relevant parameters in $\Omega_{7D}$ ocupy very small volumes compatible to a maximal Higgs boson.

The parameters grid was defined as $(m_W=[1,600])\times (m_b=[1,600-m_W])\times (m_t=[m_W+m_b,600])\times (\alpha_S(M_Z)=[0.01,0.14])$, that is it, in a region where the top quark is allowed to decay on its mass-shell to a $W$ boson and a bottom quark as discussed above. The range of $\alpha_S(m_Z)$ values were chosen in order the running quark masses could be perturbatively computed at the Higgs mass scale.
We found solutions in $\sim 0.5$\% of $10^6$ random points in $\Omega_{4D}$ which provides us with a first estimate of the probability to randomly pick one of these points. Combining this with the chance to find a maximal Higgs near the $\gamma\gamma$ peak give us a very small probability of $\sim 6\times 10^{-6}$. This, again, must be considered as an upper bound once the volume occupied by solutions in the rest of full parameters space of the SM is small.

With this probability we estimate $\Delta\approx 10^5$ as a lower bound to the amount of fine tuning, as a measure of global sensitivity, in order the Higgs mass lies at the maximum of $BR(h\to\gamma\gamma)$. We conclude that the SM is far from natural in this sense, in a similar fashion as it is concerning the radiative corrections to the Higgs boson mass. This is our main motivation to propose this principle, yet to be theoretically understood as a mechanism or a new model which could dismiss this observation as a mere astonishing coincidence.

Next we start to explore the phenomenological consequences of the proposed principle to constrain the SM and BSM models.

\section{A maximal Higgs boson principle and its phenomenological consequences}

\subsection{SM spectrum and Higgs couplings}

 If some mechanism indeed enforces the Higgs mass to maximize $BR(h\to\gamma\gamma)$, it  constrains the parameters space of the model. Suppose, for example, we have discovered the Higgs boson before the $W$ and $Z$ bosons and suppose further that we knew the maximal principle. Which vector boson masses should we expect? As is shown in Fig.~\ref{fig:mwvar}, there is a strong sensitivity against $W$ boson mass variations and only a very narrow range of $W$ masses would be compatible to the observed maximal Higgs. This would be credited as a solid prediction of the principle.

Requiring that $M_h$ be bounded in a certain region around the measured Higgs mass, how a given parameter $\theta$ would be bounded? This time we fix the SM relation between the $W$ and $Z$ masses and replace $m_Z$ and $m_c$ from $\Omega_{7D}$, in Eq.~\ref{7dparams}, by $c_f$ and $c_V$, the rescaling of the Higgs to fermions and Higgs to vector bosons couplings, respectively.

Fixing  six parameters of $\Omega_{7D}$ to their experimental values we derive the variation belt for the remaining one $\theta$. We show in table~\ref{tab2} the variation belts of all parameters in this new $\Omega_{7D}$, requiring that $M_h$ be bounded within the 99\% C.L. region around the measured mass. Except for the $\tau$ mass, which is only bounded from above by $2.4$ GeV, and the top quark mass which is still loosely bound, all the other parameters are tightly bounded around their measured present values. For example, the $W$ mass belt is obtained by varying the $W$ mass in order the central arrow in Fig.~\ref{fig:mwvar} remains inside the data point. The other belts are computed in the same way. As can be seen from table~\ref{tab2}, given the high sensitivity to the SM parameters, a maximal Higgs boson spots very small volumes of the parameters space which are consistent to all the present measured masses and couplings.

In fact, a better estimate would have to take into account the uncertainties on all the other parameters related to the branching ratios of the Higgs boson, but given the high accuracy those parameters have been measured, this simple approach suffices to show, first, the increasing level of evidence, as each SM parameter is predicted to lie in an interval which contains its observed value, and second, the potential of applications of a maximal Higgs principle.

The huge amount of Higgs data to be collected at the 13/14 TeV LHC will permit us to measure the Higgs boson mass with a very high accuracy. This will give us the opportunity to test the principle even further comparing its predictions to the observed couplings and masses related to the Higgs sector.
\begin{center}
\begingroup
\begin{table}
\begin{tabular}{c|c|c|c}
\hline
\hline
parameter & $m_W$ & $m_b$ & $m_t$  \\
\hline
variation belt & $[79.2,80.7]$ & $[4.53,4.85]$ & $[88,227]$  \\
\hline
\hline
parameter & $\alpha_S(m_Z)$ & $c_f$ & $c_V$ \\
\hline
variation belt & $[0.1165,0.1267]$ & $[0.95,1.01]$ & $[0.988,1.053]$ \\
\hline
\hline
\end{tabular}
\caption{Precision in which the parameters of the SM, $m_W$, $m_b$, $m_t$, $\alpha_S(m_Z)$, $c_f$, and $c_V$,  can be determined requiring that the maximal Higgs mass lies within the 99\% C.L. region around its measured mass. Masses in GeV.}
\label{tab2}
\end{table}
\endgroup
\end{center}

In particular, note that $c_V$ and $c_f$ are very close to unity. This indicates that couplings compatible to a maximal Higgs boson are proportional to the particles masses, that is it, originates from EWSB.

Now, let us investigate some of the features of the inner structure of the standard model.

\subsection{Number of fermion families, quark colors and the CP nature of the new boson in the SM}

\begin{figure}
\begin{center}
\includegraphics[scale=0.5]{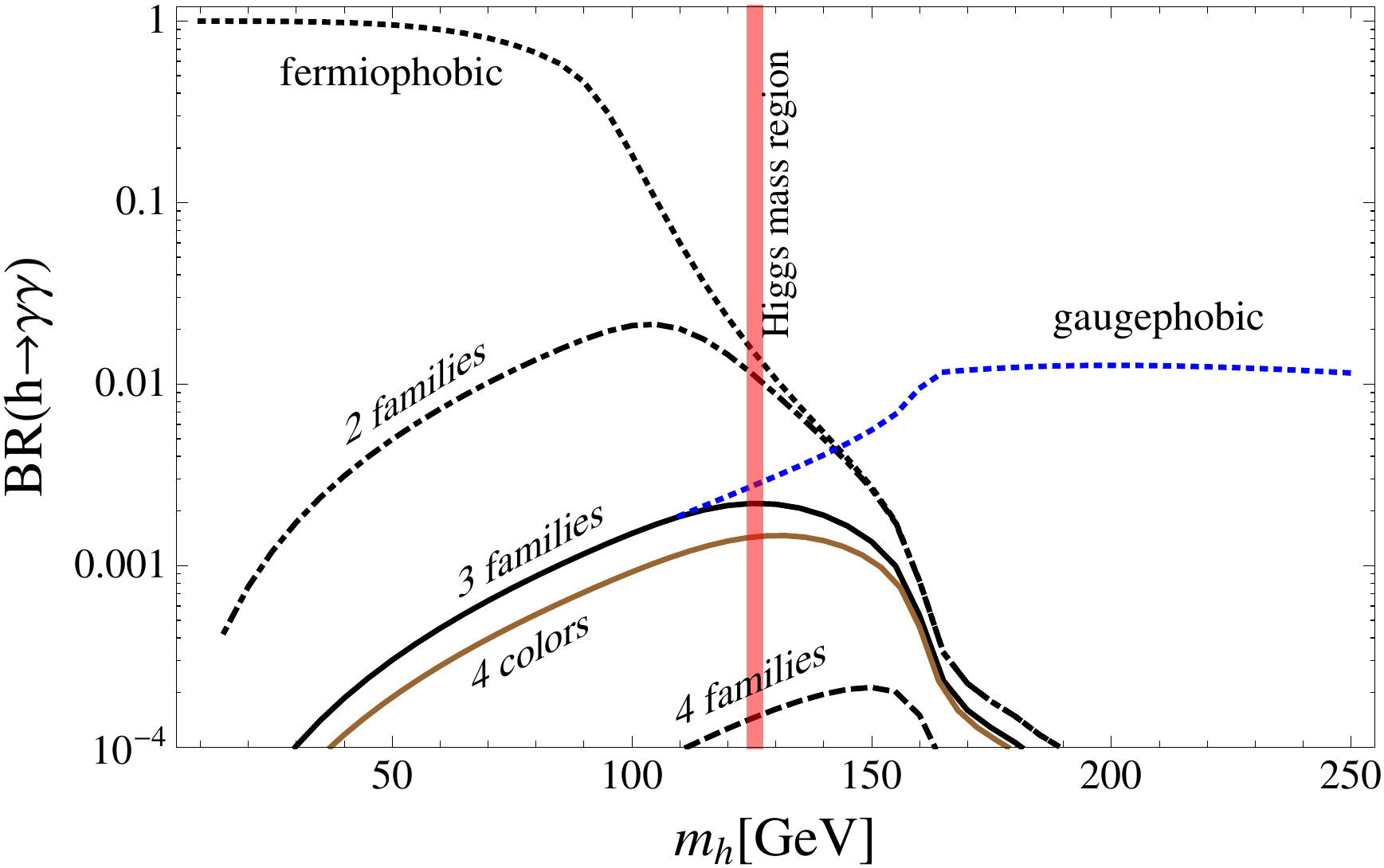}
\end{center}
\caption{Higgs branching ratio to diphotons as a function of the Higgs mass. The SM curve (3 families) and some non SM scenarios are displayed in the figure, including, a different number of fermion families, fermiophobia, gaugephobia, and a different number of quark colors. The four families curve is computed with experimentally excluded masses as we discuss in the text.}
\label{fig:familia}
\end{figure}

 Not only the masses and couplings of the SM related to the Higgs boson branching ratios are tightly bounded by the hypothesized principle, but also the inner structure of the SM. In the Fig.~\ref{fig:familia} we show the variation of the peak position as we vary the number of fermion families with all parameters fixed in their central experimental values. Only a 3 families solution is compatible with a ~125 GeV Higgs boson that maximizes $BR(h\to\gamma\gamma)$. Withdrawing the third family of fermions moves the peak to $M_h\approx 100$ GeV. On the other hand, adding a sequential fourth family with masses still not excluded by collider data, $m_{Q^\prime}=700\gev$ and $m_{L^\prime}=150\gev$, turns $BR(h\to\gamma\gamma)$ into a monotonically increasing function of $m_h$. Even for a lighter 4th family, $m_{t^\prime}=400\gev$, $m_{b^\prime}=350\gev$, and $m_{L^\prime}=100\gev$, for example, the peak position is never much smaller than 150 GeV, as shown in Fig.~\ref{fig:familia}.

The number of quark colors is also fixed by a maximal Higgs. A mass larger than 131 GeV is found if $N_c > 3$. The case of four quark colors is shown in Fig.~\ref{fig:familia}.

A fermiophobic Higgs boson is also ruled out as it does not lead to a local maximum as is shown in Fig.~\ref{fig:familia}. A Higgs boson with very weak interactions to $W,Z$ gauge bosons is another feature which is not compatible to a maximal Higgs near the observed Higgs mass value (denoted as gaugephobia in Fig.~\ref{fig:familia}).

Next, we illustrate the application of the principle to build new physics scenarios in the context of a Higgs dark portal model and the two Higgs doublet model of type-II.
\subsection{Applying the principle to a Higgs dark portal model}

Let us suppose that the Higgs is actually decaying to dark matter (DM).  The Higgs branching ratio to photons can be easily recalculated to take into account an invisible mode as $BR(h\to\gamma\gamma)=\Gamma_{\gamma\gamma}/(\Gamma_{SM}+\Gamma_{\chi\chi})$, where $\Gamma_{\chi\chi}$ is the partial width of the Higgs decay to a pair of dark matter particles. To illustrate the constraints imposed by the principle on the DM mass and its coupling to the Higgs, we choose a specific dark matter model, namely a minimal Higgs portal dark matter with the Higgs doublet $H$ and a real singlet scalar $\chi$ participating in the scalar potential of the Higgs sector~\cite{mambrini1,aoki-kanemura,Huang}
\begin{equation}
V=\frac{1}{2}m_\chi^2\chi^2+\lambda_\chi\chi^4- g_\chi|H|^2\chi^2 + V(H)
\label{eq:pot}
\end{equation}
which is symmetric under $Z_2$ transformations that render $\chi$ stable and a viable DM candidate.

After EWSB, an interaction $\chi\chi h$ arises and lead to an enhanced invisible decay rate of the Higgs boson. There is room in the parameter space of the model to accommodate the right relic abundance from WMAP and an invisible Higgs decay rate compatible with LHC8, XENON10 and CDMSII data~\cite{aoki-kanemura}. Recently, it was shown that a DM with $g_\chi\lesssim 10^{-3}$ and a mass close to $m_h/2$ is capable to explain the $\gamma$-ray spectrum at low latitudes from FERMI-LAT data~\cite{fermilat} at the same time it predicts the correct dark matter relics abundance and escape the bounds from XENON100. Moreover, the small coupling region is not ruled out by the LUX data~\cite{lux} and it will be barely allowed by the future XENON1T~\cite{okadaseto}.
\begin{figure}
\begin{center}
\includegraphics[scale=0.5]{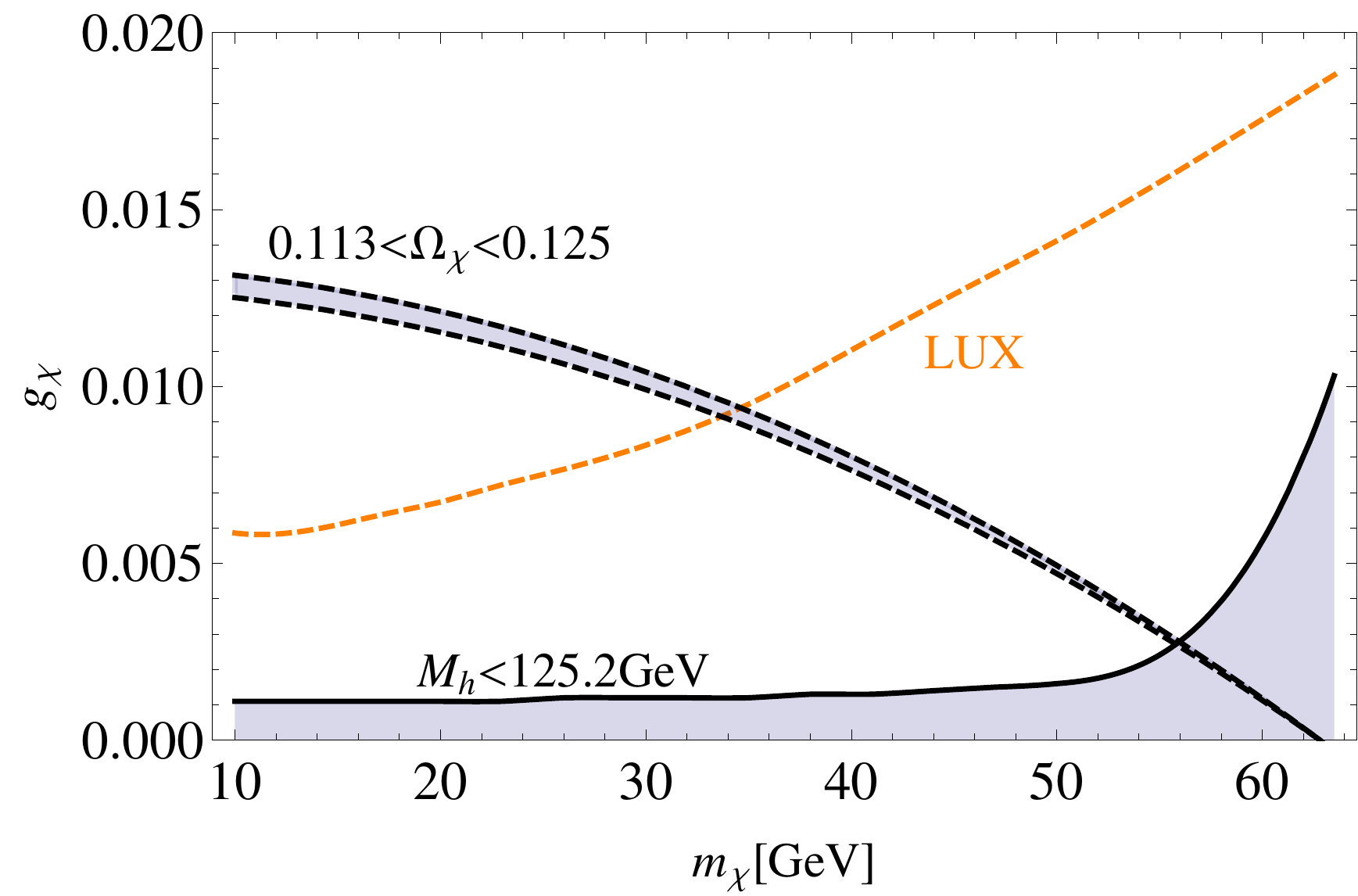}
\end{center}
\caption{Constraints on the mass and the coupling to the Higgs boson of a real scalar dark matter from a maximal Higgs principle, relic abundance and LUX data. The shaded area below the solid line represents the allowed region at 90\% C.L. from the maximal principle, the shaded area between the dashed lines is the region allowed by WMAP constraints on the DM relic abundance, and the region above the orange dashed line is excluded at 90\% C.L. by LUX~\cite{lux}.}
\label{fig:dm}
\end{figure}

In the Fig.~\ref{fig:dm} we depict the constraint from the application of the maximal Higgs principle alongside the region compatible with WMAP and the bound from LUX. Almost the entire region compatible to WMAP would be excluded at 90\% C.L. leaving a small strip for DM masses between 55 -- 63 GeV and $g_\chi < 0.01$. This region, by the way, is precisely the one cornered by the recent direct dark matter searches from XENON100 and LUX and compatible to WMAP relics abundance measurement.

Collider bounds on heavy ($\gtrsim 10\gev$) DM candidates are often less stringent than those from DM-nucleon scattering experiments, including Higgs dark portal scenarios. This is not the case whenever the DM candidate is imposed to be compatible to a maximal Higgs boson as we see in Fig.~\ref{fig:dm}. The exclusion region from the maximal Higgs boson constraint is considerably larger than the LUX region in the entire mass range allowed for a Higgs dark portal model at the 90\% confidence level.

It might be interesting to explore the following consequence of a maximal Higgs boson in a Higgs dark portal scenario. In early times soon after the Big Bang,  the temperature of the universe was high enough to produce on-shell Higgs bosons through SM particles or dark matter annihilation. In that era, about $2\times 10^{-3}$ of all Higgs bosons decayed maximally into photons, including those from dark matter annihilation. On the other hand, if the dark matter mass is such that $m_\chi > m_h/2$, even after the universe has cooled down, Higgs bosons could still be produced on-shell and decay at a maximum rate, then the dark matter to photons conversion occurred and occurs today, consequently, at the maximum possible rate as well.

\subsection{Bound on a Higgs invisible decay}

The current experimental bound on a Higgs invisible decay mode is still weak, $BR_{inv}\lesssim 0.4$~\cite{dmbound}, if we consider all the other couplings to the Higgs of strengths predicted by the SM. However, many phenomenologically interesting models predict large invisible decay branching ratios depending on the parameters of the model, as the Higgs dark portal~\cite{mambrini2} discussed in the previous section.

Instead of calculating the Higgs branching ratio to dark matter in some specific scenario, let us consider $\Gamma_{\chi\chi}$ as a free parameter and ask which values would take $M_h$ to the boundary of the given confidence belt. This can done, in a first approach, fixing all the SM parameters to their experimental values. We found the upper limit $BR(h\to\chi\chi)<1.3$\% requiring that $M_h$ lies at the 99\% C.L. region around the Higgs mass. A better estimate based on a Markov Chain Monte Carlo scan over the parameter space $\Omega_{7D}^{DM}=\{m_W,m_t,m_b,m_c,m_\tau,\alpha_S(m_Z),\Gamma_{inv}\}$ yields a very similar bound. This is a very stringent bound that may strongly constrain models where dark matter interacts to the Higgs boson.

Similar considerations were made within the ``maximum opportunity'' scenario proposed in Ref.~\cite{david}. In this case, it was found that a Higgs decaying invisibly to neutrinos would change the mass peak position to $\sim 122$ GeV.

\subsection{Implications for the type-II Two Higgs Doublet Model}

In the  Two Higgs Doublet Models~\cite{gunion}  there are two Higgs doublets that leave five physical Higgs bosons after EWSB, three neutral scalars, two CP-even ($h,H$) and one CP-odd ($A$), plus two charged states $H^\pm$.

Concerning the $h\to\gamma\gamma$ decay in the THDM~\cite{gunion},
we also have to take into account a charged Higgs loop which interferes with the $W$ and top loops. As an outcome, $BR(h\to\gamma\gamma)$ now becomes sensitive to the parameters $\alpha,\beta,m_{H^\pm}$, $m_{H^\pm}$, the neutral Higgs boson mixing angle, the ratio of the doublets {\it vevs}, and the mass of the charged Higgs bosons, respectively. Then, requiring a maximal Higgs decay to photons constrains those parameters in the same way it constrains the SM ones.
\begin{figure}
\begin{center}
\includegraphics[scale=0.6]{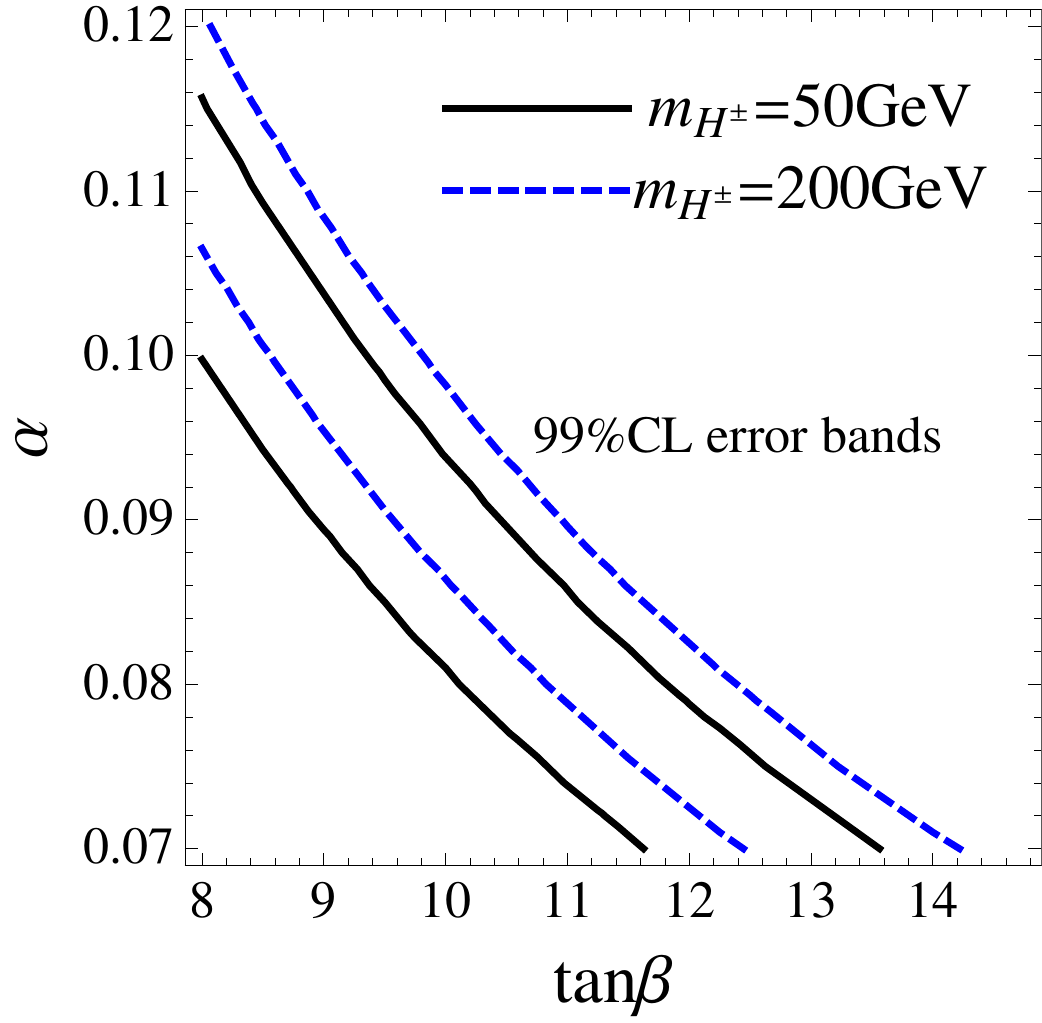}
\end{center}
\caption{The portions of the $\alpha$ {\it versus} $\tan\beta$ plane compatible to a $\sim 125$ GeV Higgs boson in a type-II THDM. The solid lines represent the 99\% C.L. limits for a charged Higgs boson of mass $50\gev$, and the dashed lines the 99\% C.L. limits for a $200\gev$ charged Higgs mass.}
\label{fig:thdm}
\end{figure}

In Fig.~(\ref{fig:thdm}) we display the 99\% C.L. bounds on the $\alpha\times\tan\beta$ plane for charged Higgs masses of 50 (solid lines) and 200 GeV (dahsed lines) of a type-II THDM. Imposing a maximum solution to $BR(h\to\gamma\gamma)$, as a function of $m_h$, constrains severely the ($\alpha,\beta$) parameters. For $\tan\beta=10$, for example, the neutral Higgses mixing angle favored by a maximal Higgs is in the very narrow interval of $[0.087,0.12]$ for a 200 GeV charged Higgs, determined by the dashed lines in Fig.~\ref{fig:thdm} , fixing the SM parameters to their experimental values. Confronting this result with those of the previous sections, a maximal Higgs is viable only as a real unmixed scalar excitation of the Higgs field.

Right after the Higgs boson discovery, the elucidation of its spin and CP assignments was addressed in order to confirm its scalar nature~\cite{spin}. Interestingly, we found no peak position in the 123--128 GeV mass range for a pseudoscalar Higgs boson of a type-II THDM~\cite{gunion}, even varying the SM parameters within their 99\% C.L. belts and varying $\tan\beta$ in a wide range.

\section{Conclusions and perspectives}
In this work, we call the attention to the notable coincidence between the measured Higgs boson mass and the Higgs mass parameter that maximizes its branching ratio to photons pairs, as predicted by the SM. We found that they are compatible within theoretical and experimental uncertainties. Even more remarkable is the fact that the peak position in $BR(h\to\gamma\gamma)$ shows a high sensitivity to the parameters and the inner structure of the SM and that only a small volume of the parameters space of the SM related to the Higgs branching ratios can produce a maximal Higgs boson. For example, slightly changing the $W$ mass lead to a peak position out of the 99\% C.L. region of the measured mass. Comparable sensitivity was found concerning several other parameters which collectively renders the SM an astonishing global sensitivity which lead us to conclude that its parameters are very unnatural from the point of view of a a precise Higgs mass that maximizes its decay rate to photons. These remarkable coincidences lead us to hypothesize the existence of a guiding principle behind the Higgs decay to photons.

Not only the SM parameters are tightly constrained by the proposed principle, but also the inner structure of the SM. Only a theory with 3 chiral fermions families, with SM couplings between a real scalar Higgs boson and the rest of the spectrum, and where the quarks carry exactly 3 color charges can lead to a maximal Higgs boson.

Any beyond the SM extension involving the Higgs sector introduce additional parameters the branching to photons might be sensible to. In order to illustrate the impact of such principle in a new physics model that affects the peak position, we investigated a minimal Higgs dark portal model and found that the principle is able to severely constrain it. Nevertheless, a model independent approach to Higgs portal scenarios showed that an invisible Higgs decay should be small, $BR(h\to \chi\chi)\lesssim 1.3$\%, in order a $\sim 125$ GeV maximal Higgs boson be plausible.


By their turn, a fourth sequential family and a THDM of type-II extension of the SM are severely constrained. A fourth family cannot produce a maximal Higgs whose mass is less than $\sim 150$ GeV, even for light new quarks and leptons. The THDM is severely constrained concerning the neutral Higgses mixing once a maximal Higgs close to the experimental value is possible only if $\alpha < 0.1$ for $\tan\beta >10$ and $m_{H^\pm}>50\gev$, what provide evidence that the maximal Higgs boson is an unmixed real scalar state.

In fact, any new SM extension affecting the Higgs sector would be constrained by such maximal principle. It would be very interesting to investigate the implications for supersymmetry models, for example, in view of the amount of fine tuning required to get the observed Higgs boson mass in those models.

The 13/14 TeV LHC will bring the opportunity to test the coincidence even further by measuring the Higgs mass with a higher accuracy. Given a better measure of the top quark mass and the Higgs couplings to the vector bosons and fermions, it will be possible to compare the peak position of $BR(h\to\gamma\gamma)$ to the measured mass and the predicted parameters to their observed values as well. Any new particle, on the other hand, should have couplings and masses in according to the principle if they are related to the Higgs sector.

A possible explanation for the coincidence that we can pursue~\cite{entropy} is that it emerges, in fact, from a maximum entropy principle. Massless quanta are the naturally best decay channels to spread the energy stored in the Higgs boson mass, increasing entropy in a maximal way compared to the massive particles.

\paragraph{\bf Acknowledgments --} A.G. Dias and E.R. Barreto thank FAPESP and CNPq for supporting this work. We also thank Oscar Eboli and Tilman Plehn for reading the manuscript.





\end{document}